\documentclass[aps,pre,onecolumn,floatfix,superscriptaddress]{revtex4}

\usepackage{graphicx}  
\usepackage{dcolumn}   
\usepackage{bm}        
\usepackage{amssymb}   
\usepackage[latin1]{inputenc}
\usepackage{amsfonts}
\usepackage{amsmath}
\usepackage{amsthm}
\usepackage{xcolor}
\usepackage{fontenc}
\usepackage{textcomp}
\usepackage{epstopdf}
\usepackage{hyperref}

\begin{document}
\title{Social adaptive behavior and oscillatory prevalence in an epidemic model on evolving random geometric graphs}
\author{Akhil Panicker}
\email{atpanicker95@gmail.com}
\affiliation{Department of Physics, Cochin University of Science and Technology, Cochin 682022, India.}
\author{V. Sasidevan}
\email{sasidevan@cusat.ac.in, sasidevan@gmail.com}
\affiliation{Department of Physics, Cochin University of Science and Technology, Cochin 682022, India.}
\date{\today}

\begin{abstract}
{\small Our recent experience with the COVID-19 pandemic amply shows that spatial effects like the mobility of agents and average interpersonal distance, together with the adaptation of agents, are very important in deciding the outcome of epidemic dynamics. Structural and dynamical aspects of random geometric graphs are widely employed in describing processes with a spatial dependence, such as the spread of an airborne disease. In this work, we investigate the interplay between spatial factors, such as agent mobility and average interpersonal distance, and the adaptive responses of individuals to an ongoing epidemic within the framework of random geometric graphs. We show that such spatial factors, together with the adaptive behavior of the agents in response to the prevailing level of global epidemic, can give rise to oscillatory prevalence even with the classical SIR framework. We characterize in detail the effects of social adaptation and mobility of agents on the disease dynamics and obtain the threshold values. We also study the effects of delayed adaptive response of agents on epidemic dynamics. We show that a delay in executing non-pharmaceutical spatial mitigation strategies can amplify oscillatory prevalence tendencies and can have non-linear effects on peak prevalence. This underscores the importance of early implementation of adaptive strategies coupled with the dissemination of real-time prevalence information to manage and control the epidemic effectively.}
\end{abstract}
\maketitle

\section{Introduction}
\label{sec0}
Epidemics have always been a threat to humanity since ancient times. The black 
death wiped out two-thirds of the European population in the $14^{th}$ century 
\cite{blackdeath}. COVID-19 has so far caused more than 6.9 
million deaths \cite{coronadata}. Understanding and controlling such events is, therefore, of paramount importance for our own survival on Earth. The dynamics of 
epidemics have been analyzed using various types of mathematical and 
computational models. Such models are of immense importance as they can give us 
quantitative insights into the dynamic process of an epidemic. Together with the 
knowledge generated in various other disciplines and field data,  models help us  
make informed decisions to deal effectively with a pandemic. The information gained 
from models of epidemics that incorporate pharmaceutical and non-pharmaceutical 
interventions are important in order to have better control over the epidemic 
\cite{Ahmed2021,Gatto2020,Lang2018,Colizza2007}.

The major type of mathematical model of the epidemic is the compartmental model in 
which a population is divided into various compartments such as S (Susceptible), 
I (Infected), R (Recovered), etc, based on the state of infection of individuals. 
In the simplest setting, such models constitute a set of rate equations for the 
fraction of individuals in various compartments and are mean-field in nature 
\cite{Ottar2018,Keeling2008,Rothman2008}. Real-world population structures are different from 
the ones typically considered in the mean-field equations of compartmental models. In a 
more realistic setting, population structure is modeled as a network in which 
individuals are the nodes and connections are the links of a complex network. 
In the network, two individuals are assumed to be `connected' if the disease can 
be transmitted between them. In addition to such direct physical connections, one may also consider other types of connections between the agents, like that on a social network, which will affect the disease propagation. Models of epidemic spread on such topological (and often multilayer) networks have been extensively investigated in the past 
\cite{Turner2020,Pastor2015,Cohen2010,Junfen22,Sun2022,Shuofan2023}. 

In many real-world settings, spatial factors such as the average distance 
between individuals and their mobility play a crucial part in deciding the 
structure of a contact network and will influence any dynamic process defined on 
such a network. The networks where the connectivity is decided by a 
distance-dependent measure are called random geometric graphs \cite{Barthelemy2011,Penrose2007}. Such 
spatial factors, which are normally not considered in epidemic models on 
topological networks, have gained increased recent attention in the wake of the 
COVID19 pandemic 
\cite{chang2021,loring2020,wong2020,pujari2020,Melin2020,Kang2020}.  In such 
spatial network models of the epidemic, individuals or nodes are embedded in 2D 
space in which 
connections exist between two nodes only if they are closer than a 
characteristic distance or the transmission range of a disease. The value of the 
characteristic distance can vary from zero for a disease that transmits only by 
person-to-person direct contact up to several meters for airborne diseases. 
The characteristic distance may also depend upon certain preventive 
strategies adopted by individuals, such as mask usage. So, the structure of the contact network, in general, will be dynamic as well as disease-dependent. 
Thus, models of epidemics on spatial networks can give us valuable insights into 
the dynamics of a disease in a population by incorporating factors like mobility 
of individuals and other adaptive spatial intervention strategies like social-distancing. Such considerations become all the more important when preventive measures like vaccination are not available. Recent research shifted focus from traditional approaches to considering individual-level behavior and the response of agents to an ongoing epidemic, and detailed contact patterns in populations. These studies span the developmental trajectory of theoretical epidemiology, evolving from classical mean-field models to recent spatially aware frameworks, often rooted in statistical physics \cite{WangStati,Markosocial}. 

Various pharmaceutical and non-pharmaceutical intervention strategies can be 
employed to control an epidemic. For an air-born disease like COVID-19, mask 
usage, social distancing, and mobility restrictions are some of the most 
important non-pharmaceutical techniques that can be used to control the 
epidemic. Such intervention actions will have a direct bearing on the contact 
structure of the population 
\cite{Paulo2022,Arthur2021,Havlin2020,Lopez2020,Vrugt2020,Maier2020,Eli2011,
Funk2009, Caley2008}. Prior studies have explored the impact of community lockdowns and travel restrictions on epidemic-spreading dynamics. Investigations reveal that the effectiveness of community lockdowns relies on the near-complete closure of links outside the communities \cite{Mgosac21}.  Mask usage  
will reduce the `connectivity' of the network by reducing the transmission range 
of viral particles between persons. Social distancing and mobility restrictions 
will also reduce the connectivity or the mean degree of the network by keeping 
individuals apart, as in a low-density population. Since the effect of all such adaptive
intervention actions are effectively the same, viz, reduction of connectivity of 
the network, we will refer to all such actions by the generic term `Social 
Adaptation' (SA). Previous works that incorporate similar social adaptation 
have shown that oscillations in prevalence can arise due to 
individual payoff-based game-theoretic considerations by the agents 
\cite{Khazaei2021,Jianping2021,Liu2021,Glaubitz2020,Just2018}.     

In this work, we investigate how the adoption of such non-pharmaceutical adaptive
intervention strategies by the agents who are spatially distributed and are 
mobile, affects the outcome of SIR dynamics. The connectivity structure of 
agents is modeled by random geometric graphs, which evolve by the adaptive 
actions of individuals as well as their mobility. The adaptive action of agents 
is incorporated via a threshold model for social adaptation i.e. their decision 
to follow SA depends upon the level of global prevalence with respect to a 
threshold prevalence.  We show that such adaptive actions by the agents can give 
rise to oscillations in the prevalence of the disease even with simple SIR 
dynamics. We quantitatively characterize the effectiveness of 
non-pharmaceutical adaptive intervention strategies in controlling the epidemic. We 
obtain conditions under which effective reduction in the peak prevalence can be 
obtained from numerical solutions as well as simulations. We also study the 
effect of delays in executing such non-pharmaceutical threshold-based SA strategies on the 
epidemic. In this case, we show that such delays accentuate oscillations in the 
prevalence and have a non-linear effect on the peak prevalence. Our study 
shows how spatial factors like mobility and average interpersonal distance, together with the adaptive actions of the population - either voluntary or 
enforced- can give rise to epidemic waves in time. 

We also conduct a comparative analysis of our proposed model with real-world COVID-19 data from different countries. Specifically, we show that a double threshold model where agents employ social distancing when the prevalence goes above a particular level and relax the restrictions when the prevalence goes below a level can describe the oscillations seen during the pandemic. We would like to emphasize that, while our idealized model can capture some aspects of the observed data, it is neither intended nor expected to capture all the complexities of a real-world pandemic. It should be viewed as a model with special emphasis put on the aspect of spatial adaptation behavior and its effect on disease dynamics.

The paper is organized as follows. In Sec.~\ref{sec2}, we introduce the SIR 
model on evolving random geometric graphs and discuss its threshold behavior. 
We characterize the effect of the mobility of agents on the SIR dynamics. In 
Sec.~\ref{sec3}, we discuss the effect of non-pharmaceutical adaptive 
strategies by the agents on the dynamics of the epidemic and how that leads to 
oscillations in the prevalence. In Sec.~\ref{sec4}, we consider the effects of 
delays in implementing the adaptive strategies and compare it with real-world COVID-19 data in Sec.~\ref{sec5}, followed by a discussion of our results in Sec.~\ref{sec6}. 

\section{SIR dynamics on evolving Random Geometric Graphs}
\label{sec2}
\begin{figure}[!htb]
\includegraphics[width=0.95\linewidth]{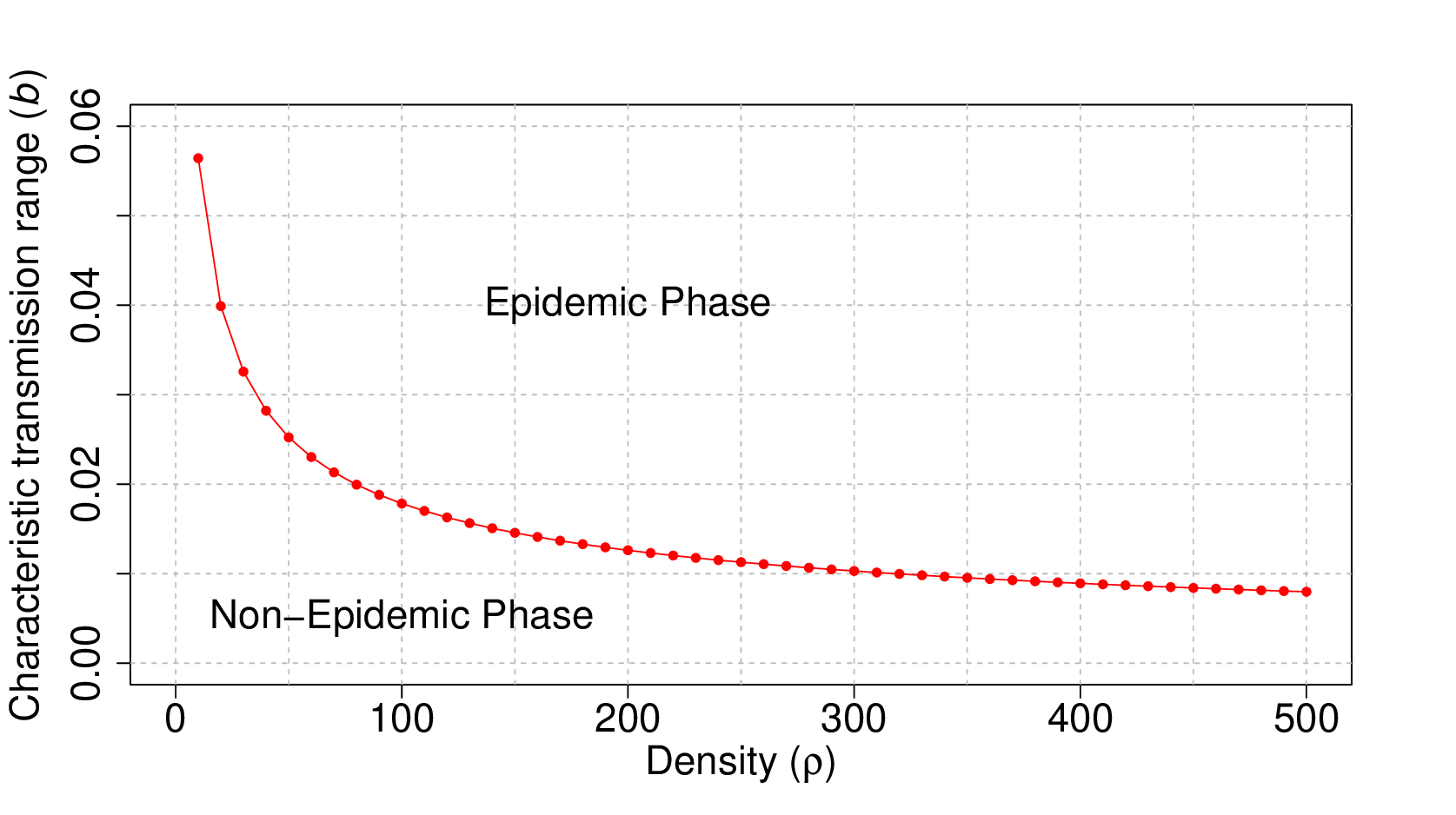}
\caption{{\bf Variation of critical characteristic transmission range with density.}
The critical characteristic range is calculated using Eq.~\ref{eqn:eq7} for different values of density. Throughout the present work, we will 
use the values transmission probability $(\beta)$ = 0.5, recovery rate 
($\gamma$) = 0.05, and density of agents $\rho = 500$. An epidemic outbreak cannot happen in the region below the curve, whereas it is allowed above it.}
\label{fig1}
\end{figure}

We will follow the works of \cite{Glaubitz2020,Peng2019,Buscarino2008} in 
defining an 
epidemic model with spatially distributed agents. We consider a spatial network 
in which $N$ individuals (nodes) are distributed uniformly and randomly in a 
square patch of length $L$ with density $\rho = \frac{N}{L^{2}}$.  Two nodes are assumed to be `connected' and can potentially pass on the disease if they are closer than a characteristic 
transmission range $b$. At each time step, an agent moves from its current 
location and assumes a new random position within a circular patch of radius 
$m_0$ with the current location as the center.  This will lead to a new 
spatial connectivity structure at each time step. We call $m_0$ as the mobility parameter. SIR dynamics is implemented on this evolving RGG 
where Susceptible (S), Infected (I), and Recovered (R) are the 
compartments.  When $m_0 = 0$, the nodes are static. When $m_0 \sim L$, over time, all the individuals interact with all others. These are the extreme cases of mobility. In addition to this, if we 
assume that the change in the connectivity structure of the network and the epidemic process happens at the same rate, we can write down mean-field equations 
to model the process. Let $\beta$ be the probability with which infection 
is transmitted to a neighbor of an infected individual, and $\gamma$ be the probability that an infected individual recovers from infection at any time 
step. 
At any time step $t$, the number of Susceptible, Infected, and Recovered agents are denoted by $S(t)$, $I(t)$, $R(t)$ such that  
\begin{equation}
	S(t)+I(t)+R(t)=N
\end{equation}
or
\begin{equation}
	s(t)+i(t)+r(t)=1
\end{equation}
Where $s(t) = S(t)/N$, $i(t) = I(t)/N$, $r(t) = R(t)/N$ are the normalized 
values of the number of Susceptible, Infected, and Recovered agents, respectively. 
Now, the probability of a susceptible agent not being infected by any of its 
infectious neighbors in a given time step is  $(1-\beta)^{n}$ where 
$n = \rho \pi b^{2} i(t) $ is the average number of infected neighbors 
inside a disk of radius $b$.
Therefore, the equations for the evolution of the fraction of agents in 
different compartments take the form, 
\begin{equation}
	s(t+1)= s(t) - s(t)[1-(1-\beta)^{\pi b^{2} \rho i(t)}]\label{eqn:eq1}
\end{equation}
\begin{equation}
	i(t +1)= i(t)-\gamma i(t)+s(t)[1-(1-\beta)^{\pi b^{2}\rho i(t)}]\label{eqn:eq2}
\end{equation}
\begin{equation}
	r(t+1)= r(t)+\gamma i(t)\label{eqn:eq3}
\end{equation}
For small values of $\beta$, Eq.~\ref{eqn:eq2} becomes,
\begin{equation}
	i(t+1) \approx i(t)-\gamma i(t)+[1-i(t)-r(t)]\beta{\pi b^{2}\rho i(t)}\label{eqn:eq4}
\end{equation}
Since the recovered compartment $r(t)$ will be very small at the beginning of an 
epidemic, letting $r(t) \to 0$, Eq.~\ref{eqn:eq4} becomes, 
\begin{equation}
	i(t+1) \approx i(t)-\gamma i(t)+[1-i(t)]\beta{\pi b^{2}\rho i(t)}\label{eqn:eq5}
\end{equation}
Therefore, for the epidemic to grow, we must have,
\begin{equation}
	1-\gamma + \beta{\pi b^{2}\rho} \geq 1 \label{eqn:eq6}
\end{equation}
Thus, for a given density, the critical characteristic transmission range for an outbreak to happen is given by 
\begin{equation}
	b_{epi} = \sqrt{\frac{\gamma}{\rho \pi \beta}}\label{eqn:eq7}
\end{equation}
For values of $b$ above $b_{epi}$, epidemic outbreak happens and below it,  
epidemic cannot happen \cite{Peng2019,Buscarino2008}.  It is instructive to 
compare the above critical transmission range with the condition for the 
formation of a giant connected component in a continuum percolation problem
of overlapping discs with radius $b$. In the latter, overlapping discs of 
radius $b$ are randomly distributed in a plane with density $\rho$. When the 
value of the radius $b$ is sufficiently high, a giant connected component forms 
in the system, signaling a phase transition. 
Denoting the critical radius of discs at which the transition occurs by
$b_{gc}$, we know that \cite{Mertens2012}
\begin{equation}
b_{gc} \approx \sqrt{\frac{1.128}{\pi \rho}}\label{eqn:eqn8}
\end{equation}  
When the radius is below the above critical value, no large connected 
component exists in the system. It is clear that, in a population with no 
mobility, $b_{gc}$ will act as a lower threshold value 
of the characteristic transmission range below which no epidemic spread can occur. However, 
when there is mobility, the lower threshold is given by Eq.~\ref{eqn:eq7}. Therefore, we 
have the relation
\begin{equation}
	b_{epi} \approx b_{gc}\sqrt{\frac{\gamma}{1.128 \beta}}\label{eqn:eq9}
\end{equation}
Fig.\ref{fig1} shows the variation of the critical characteristic range 
$b_{epi}$ with the density of the population. For a given value of $\beta$ and 
$\gamma$, such a curve demarcates epidemic and non-epidemic regions. 

Relaxing the assumptions about either the mobility or the 
rate of the two processes will require explicit consideration of the network 
structure. We use Monte Carlo simulations in these cases to obtain the results. 
Especially we will consider the two extremes of mobility i.e. the cases of static agents ($m_0 = 0$) and fully mobile agents $m_0 \sim L$.

\begin{figure}[!htb]
	\includegraphics[width=0.95\linewidth]{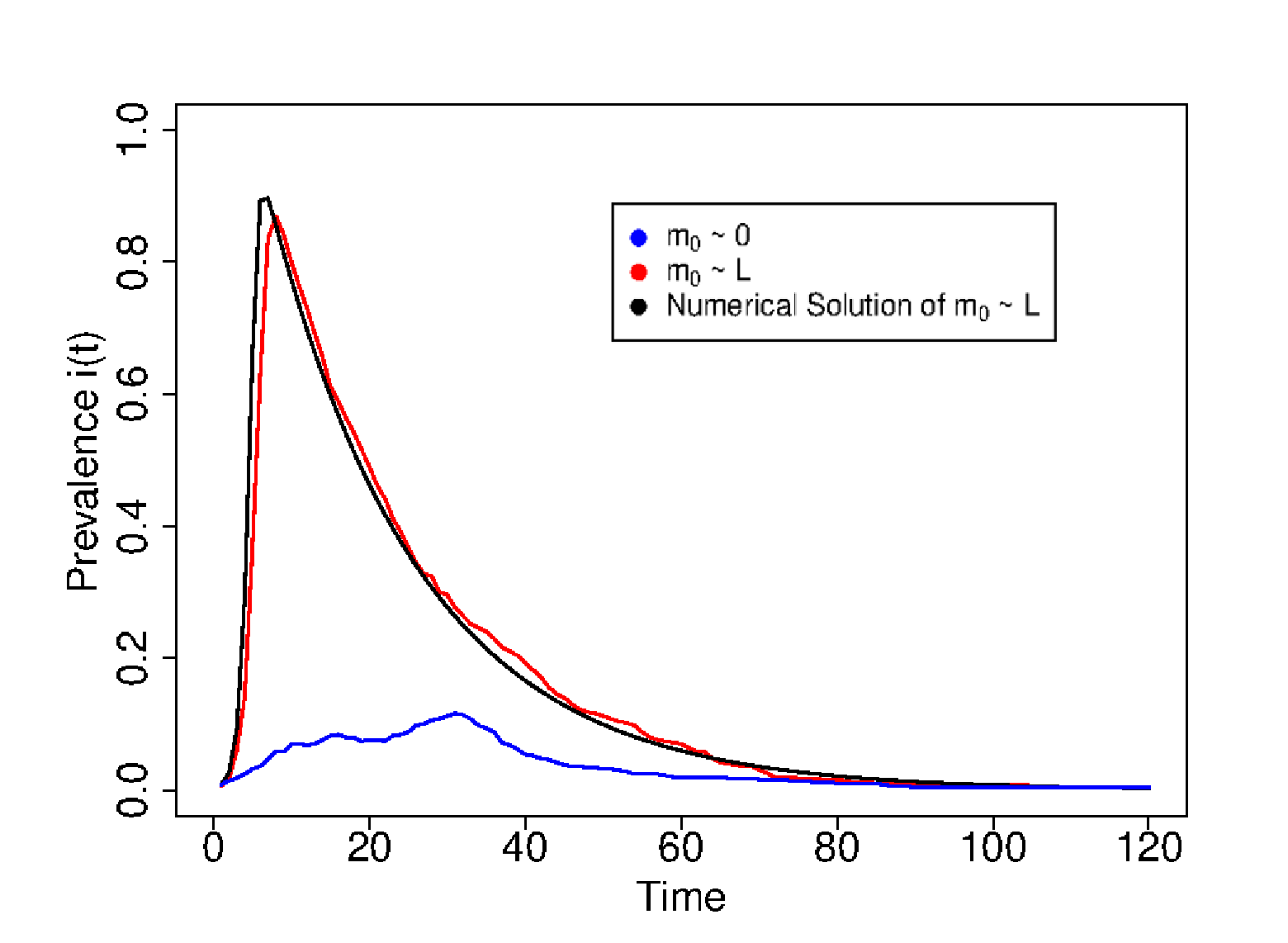}
	\caption{{\bf Variation of prevalence with time.}
    Prevalence $i(t)$ is plotted for numerical and simulated solutions of an epidemic with mobility $m_0 \sim L$. The simulation result is also shown for the case where the agents are not mobile $m_0 = 0$. Characteristic transmission range  $b = 0.05$.  The cases $m_0 = 0$ and $m_0 \sim L$ act as two extreme scenarios. The agents are spatially static in the former, whereas they get fully mixed over time in the latter.}
	\label{fig2}
\end{figure}

Fig.~\ref{fig2} shows the prevalence over time curves for the cases with and 
without mobility of agents. In the present work, we will use the values, 
transmission probability $\beta = 0.5$  
and recovery rate $\gamma = 0.05$. This means that there is a $50\%$ chance 
that a susceptible person who is within the characteristic range of an infected 
person will get the disease, and the average number of days for recovery is 20. 
Changing these values does not affect the qualitative nature of the results.  
From the figure, we can see that the mobility of agents has a pronounced effect 
on both the peak prevalence and the duration of the epidemic. We can see that 
the peak prevalence more than doubled when the agents are fully mobile.  

\begin{figure}[!htb]
  \includegraphics[width=0.95\linewidth]{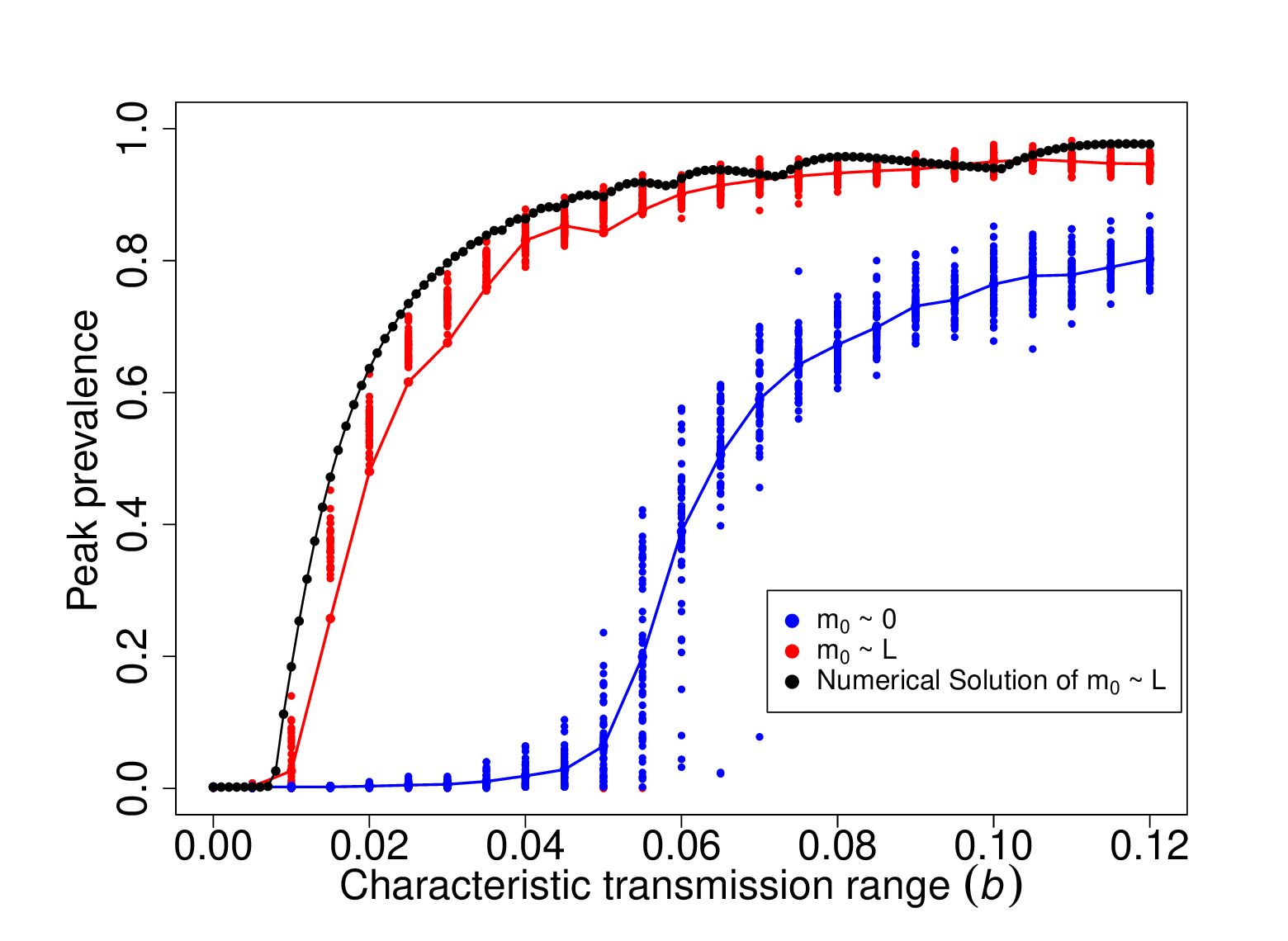}
	\caption{{\bf Variation of peak prevalence with characteristic transmission range.}
		Comparison of peak prevalence with characteristic transmission range $b$ is 
shown for the extreme scenarios of $m_0 = 0$ and $m_0 \sim L$.  
With static agents ($m_0 = 0$), a much larger transmission range is required for peak prevalence to have significant values.}
	\label{fig3}
\end{figure}

Note that for a given disease, both $\beta$ and $\gamma$ are fixed quantities 
over which we do not have any control in general. Two controllable 
parameters here are the characteristic range $b$ and the density of agents 
$\rho$ (Another potentially controllable parameter is $m_0$). Characteristic 
range $b$ may be altered by measures such as mask usage, while $\rho$ may be altered by measures such as social distancing or lock-downs. 
Note that a change in $b$ can also be viewed as a corresponding change in the 
density $\rho$. Fig.~\ref{fig3} shows the variation of peak prevalence with the 
characteristic range. Again, we compare the results with the case in which there 
is no mobility ($m_0 = 0$). The cases $m_0 = 0$ and $m_0 \sim L$ act as two
extreme scenarios, and we anticipate an intermediate behavior in the case of a 
population with in-between values for the mobility parameter $m_0$.

\section{Effect of threshold-based adaptation strategies on the epidemic}
\label{sec3}
An effective non-pharmaceutical intervention strategy to contain a disease like 
COVID-19, which can be transmitted from person to person via air, is to reduce the 
average effective interpersonal distance in a population. Measures such 
as mask usage, promoting social distancing, or partial or complete lock-downs 
are all examples of such adaptive strategies. Such measures could be either self-imposed 
by the agents or imposed by an external agency. Such measures are usually 
imposed and removed depending on the prevalence of the disease in the population 
although this may not be the sole criteria based on which such decisions are 
made.  Ideally, we would like such strategies to have the 
effect that the average Euclidean distance between the individuals in the 
population becomes greater than the characteristic transmission range of the 
disease.  For a given disease, we may view non-pharmaceutical intervention 
strategies as either {\bf(a)} Increase the average distance between the 
agents or {\bf (b)} Reduce the transmission range of the disease $b$. The first 
method can be implemented by assuming that the length of the system is increased 
by a factor of $f$ while keeping the number of agents the same when the agents 
follow social adaptation such that the mean distance between individuals increases 
by a factor of $f$ where $f \geq 1$. So the density changes from $\rho = 
\frac{N}{L^{2}}$ to $\rho^{SA} = \frac{N}{(Lf)^{2}}$ where $\rho^{SA}$ is the 
density of the agents while following social-adaptation strategies. The second 
method can be implemented by assuming that the characteristic transmission range 
$b$ is reduced by a factor of $1/f$. While both methods are mathematically the same, 
the latter describes situations like using face masks, which effectively reduce 
the transmission range of the disease. Here, we will employ a reduction of 
$b$ to implement SA and will call $f$ as the SA factor.    

\begin{figure}[!htb]
	  \includegraphics[width=0.95\linewidth]{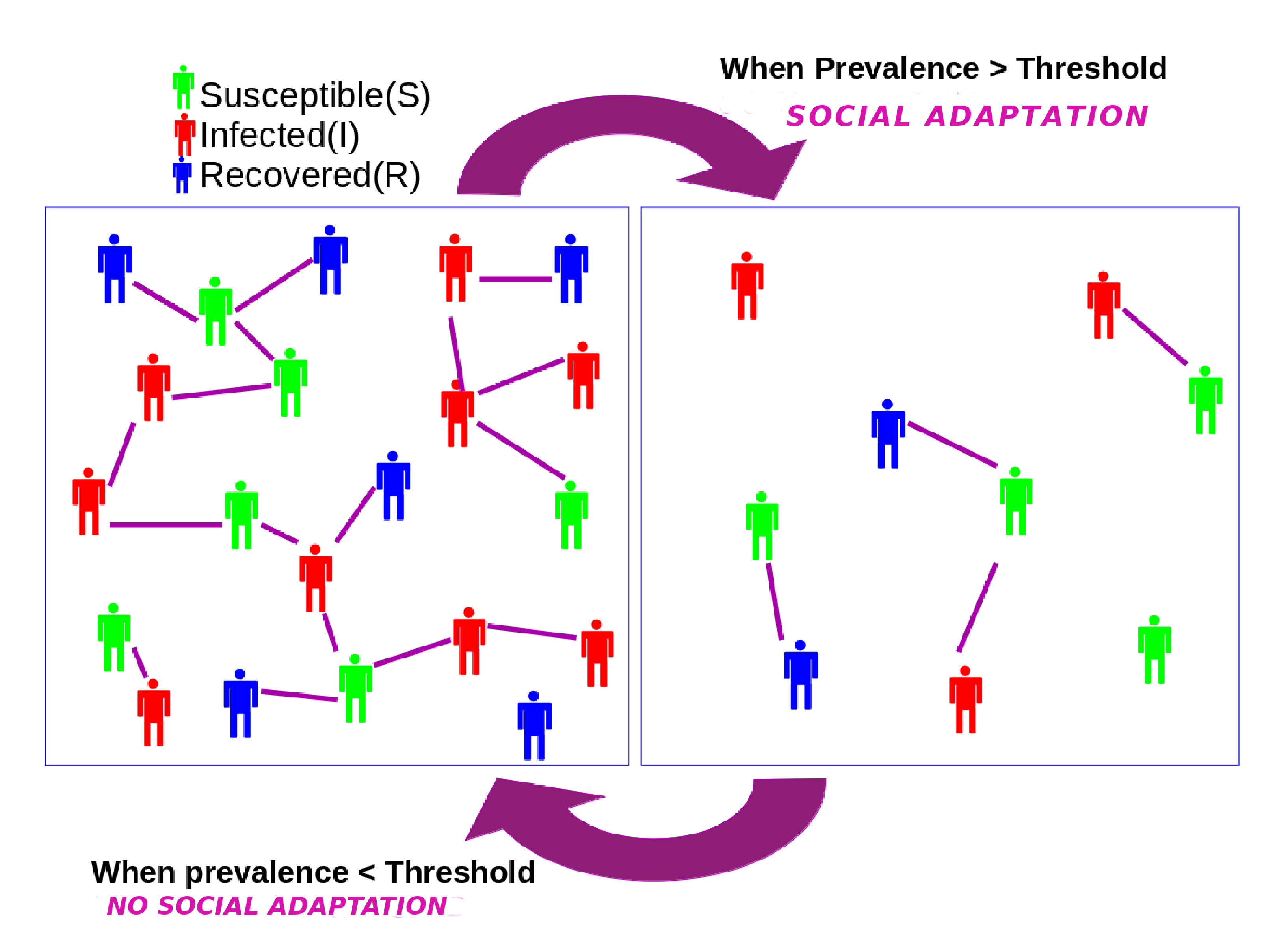}
	\caption{{\bf Schematic representation of the adaptive behavior of agents in the model}. Agents are distributed spatially in a 2D square patch of size $L \times L$ with a density $\rho$ and move around. Agents employ social adaptation whenever the global prevalence of the disease is greater than a threshold $i_c$ and discard it otherwise. Adaptation results in larger interpersonal distance between the agents, which is equivalent to a reduction in the density of the agents (see text).}
	\label{fig5}
\end{figure}

We assume that agents or a central agency monitor the level of global 
prevalence in the population. Whenever the epidemic prevalence goes above a 
pre-defined threshold value $i_c$, agents follow social distancing from the 
next time step till the prevalence is reduced below the threshold (See 
Fig.~\ref{fig5}). The characteristic transmission range $b$ then evolves 
according to,  
\begin{equation}
	b(t+1)=\left\{
	\begin{array}{@{}ll@{}}
		\ \frac{b(t)}{f}, & {if}\ i(t)>i_c \\\label{eqn:eq10}		
b(t), & {otherwise}
	\end{array}\right.
\end{equation}

where $b(0)$ is the original characteristic transmission range of the disease in the absence 
of any SA. The mean degree of the network thus assumes either of the  two values 
$\rho \pi b^{2}$ and  $\frac{1}{f^{2}}\rho \pi b^{2}$ depending upon the 
prevalence at any time step.  

\begin{figure}[!htb]
	  \includegraphics[width=0.95\linewidth]{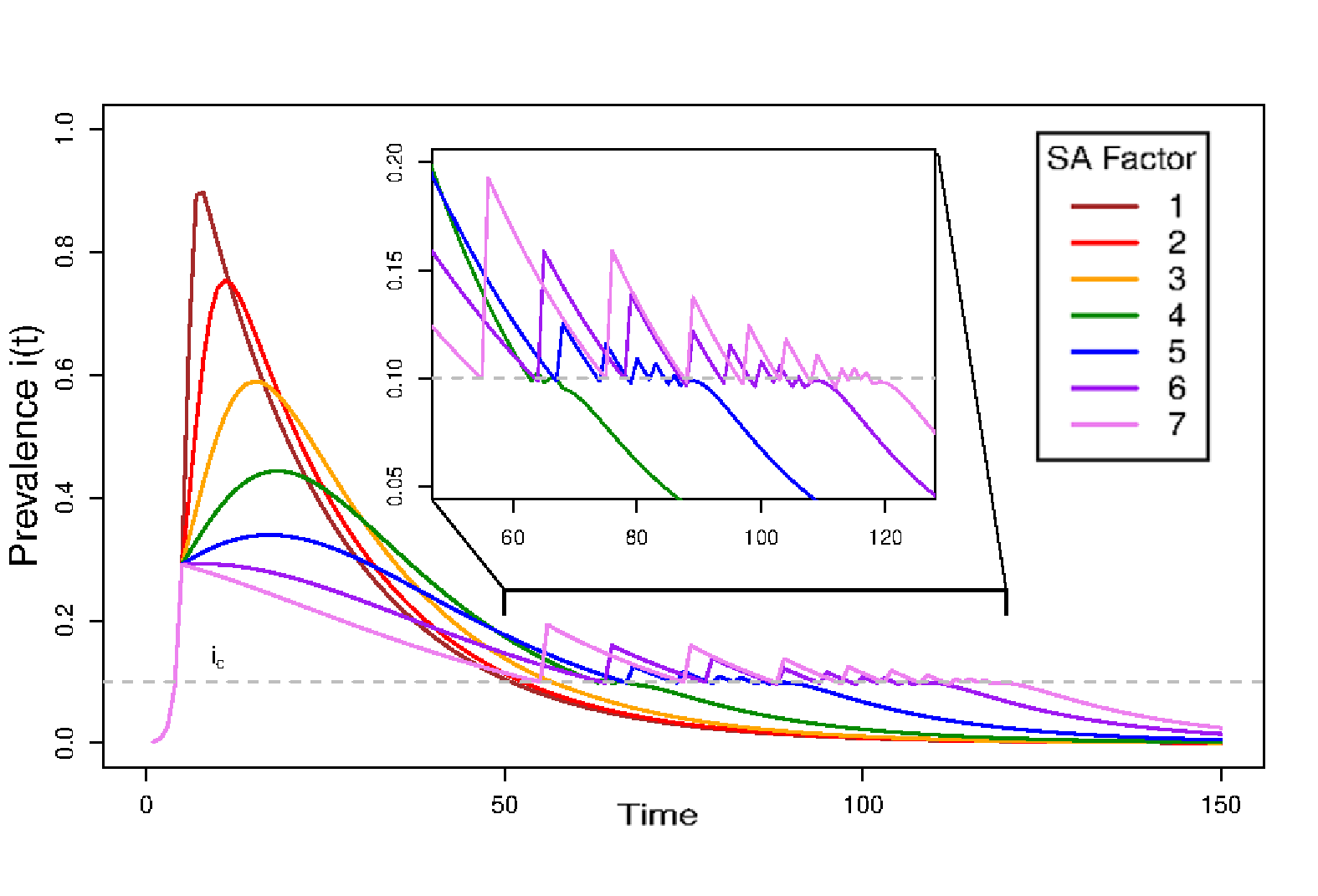}
	\caption{{\bf Numerical solution of prevalence as a function of time for 
various values of social adaptation factor $f$ with mobility $m_0 \sim L$.} $f = 1$ corresponds to the case with no adaptive measures employed. $b(0) = 0.05$. Beyond a specific large value of SA factor, prevalence drops significantly, and oscillations develop.}
	\label{fig6}
\end{figure}

We will first consider the situation of $m_0 \sim L$. In this case, Fig.~\ref{fig6}  gives a comparison of the prevalence with and without SA. As 
we increase the SA factor $f$, the peak prevalence continues to drop, but a significant drop in the peak prevalence is achieved only beyond a critical value 
of $f$ ($f \approx 6$ in the figure). This can be understood based on the fact 
that for lower values of $f$, there is still an effective giant cluster in the 
system aiding the epidemic in spreading. In other words, SA is not enough to bring 
the system below the critical line in Fig.~\ref{fig1}.  As the value of $f$ goes 
beyond the critical value, we can see that the prevalence oscillates around the 
threshold value $i_c$, which indicates that the characteristic transmission 
range went below its critical value. The threshold value of $f$, say $f_{th}$ is 
related to the critical value of the characteristic transmission range 
$b_{epi}$ by

\begin{equation}
	f_{th} = b_{epi} \sqrt{\frac{\rho \pi \beta}{\gamma}}\label{eqn:eq11}
\end{equation}

\begin{figure}[!htb]
	 \includegraphics[width=0.95\linewidth]{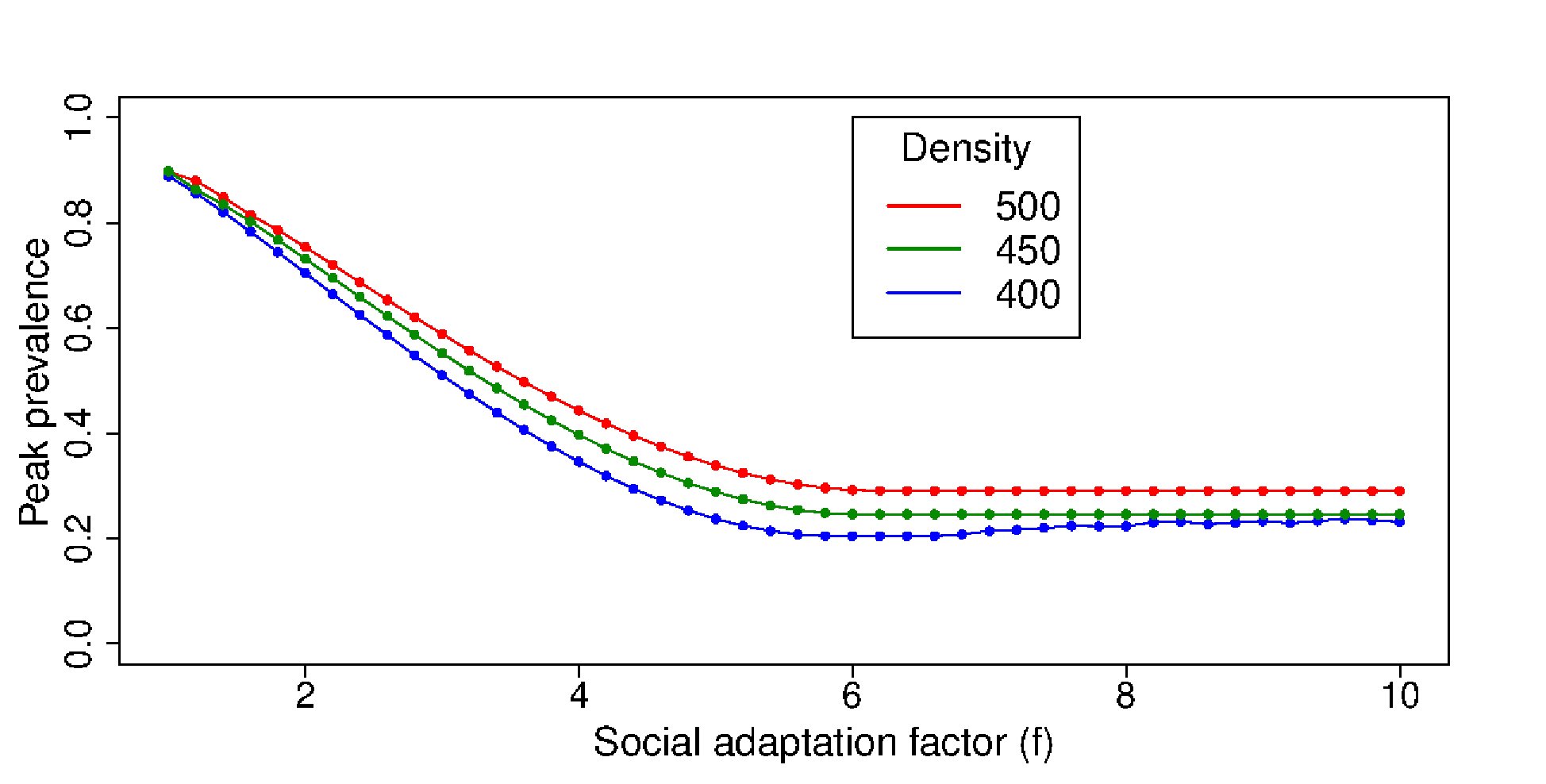}
	\caption{{\bf Numerical solution for variation of peak prevalence with 
social adaptation factor $f$ for various values of densities $\rho$}. $m_0 \sim L$, $b(0) = 0.05$ and threshold $i_c= 0.1$. Beyond a critical value of $f$, social adaptation is ineffective in reducing peak prevalence further.}
	\label{fig7}
\end{figure}

Fig.~\ref{fig7}  shows the variation of peak prevalence with the SA factor $f$. As we increase the value of $f$, peak prevalence reduces till the critical value of $f$, and thereafter the peak prevalence stagnates. A further increase of $f$ is not effective in reducing the peak prevalence and is not optimal from a socio-economic point of view as it imposes additional restrictions on the population without any additional benefits. 

\begin{figure}[!htb]
	  \includegraphics[width=0.95\linewidth]{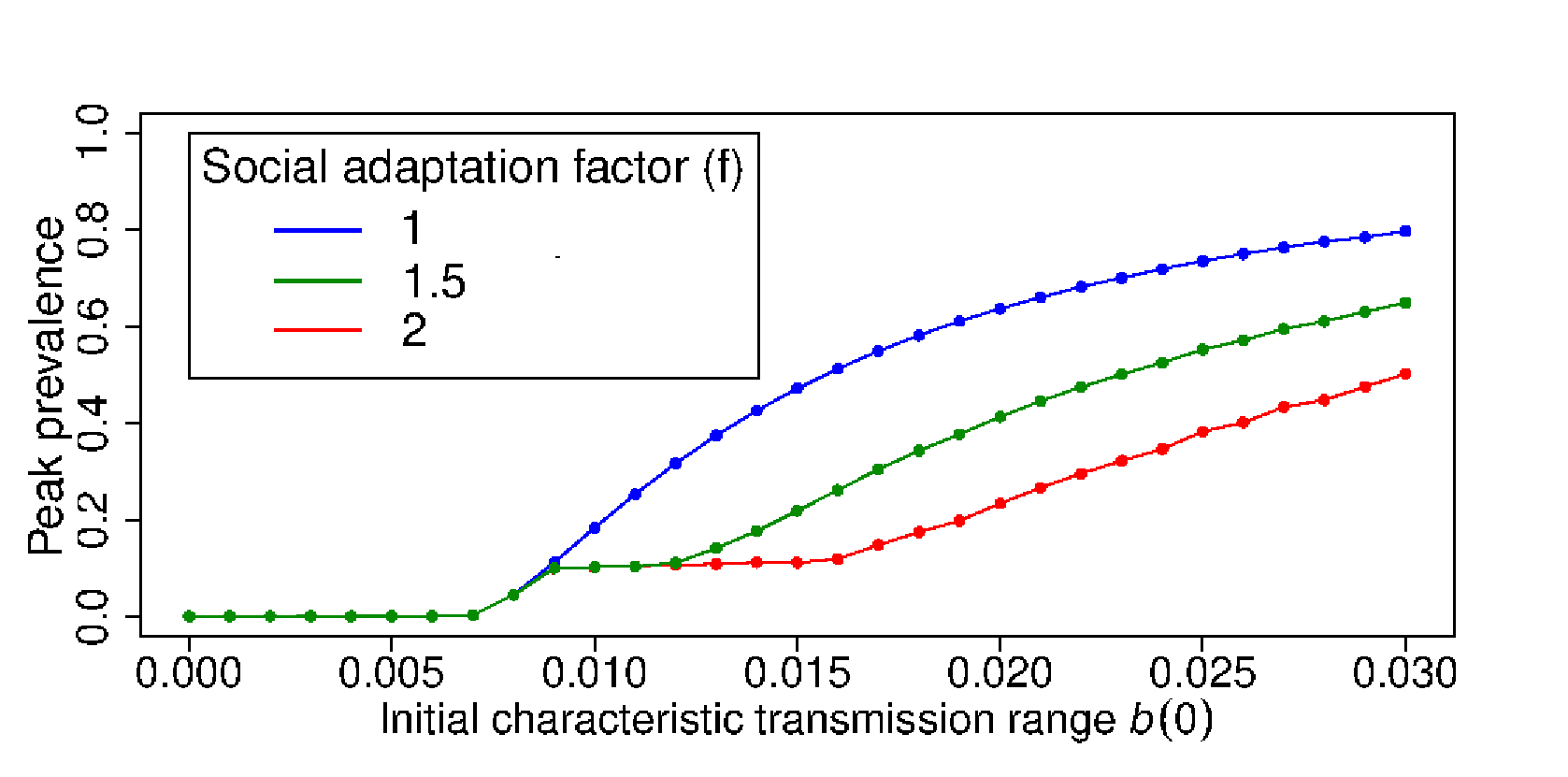}
	\caption{{\bf Numerical solution for variation of peak prevalence with 
initial characteristic transmission range $b(0)$ for various values of social 
adaptation factor 
$f$.} $m_0\sim L$ and $i_c= 0.1$. Different levels of social adaptation become effective in controlling peak prevalence only for a range of values of $b(0)$.}
	\label{fig8}
\end{figure}

It is instructive to look at the peak prevalence as a function of the initial 
characteristic transmission range $b(0)$ for different values of the social 
distancing factor $f$, which is shown in Fig.~\ref{fig8}. We can see that the 
peak prevalence becomes non-zero above the critical threshold given by 
Eq.~\ref{eqn:eq7}. However, for a particular value of $f$, the peak prevalence is 
contained at the threshold value $i_c$ for a range of values of $b(0)$.  As we 
further increase $b(0)$, the adaptation is no longer effective in controlling 
the epidemic, and the peak prevalence again rises after a specific value of 
$b(0)$. For higher values of $f$, the range over which the peak prevalence 
remains at the threshold value is also higher. The behavior can be understood 
based on the critical characteristic range $b_{epi}$ given in 
Eq.~\ref{eqn:eq7}. The peak prevalence is contained at the threshold $i_{c}$ only when the adaptation 
brings the effective interpersonal distance to values below $b_{epi}$. 

\begin{figure}[!htb]
	  \includegraphics[width=0.95\linewidth]{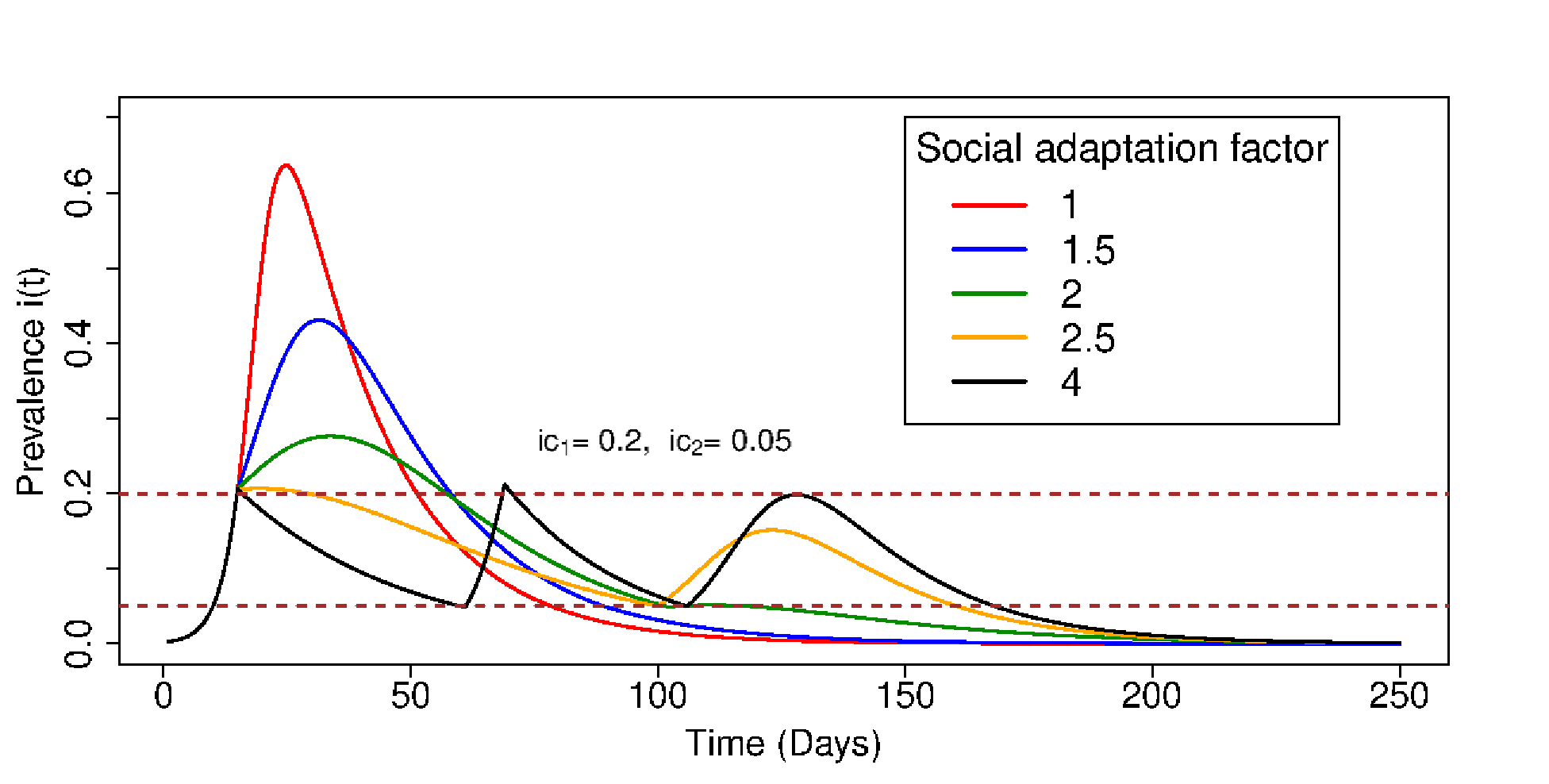}
	\caption{{\bf Numerical solution of prevalence as a function of time for 
various values of social adaptation factor $f$ for the double threshold model.} $b(0) = 0.02$ and $m_0 \sim L$. $i_{c1} = 0.2$ is the upper 
threshold and $i_{c2} = 0.05$ is the lower threshold. Beyond a specific value of SA factor, prevalence drops significantly, and oscillations develop.}
	\label{doublecutoff}
\end{figure}

We further extend the model to include a lower threshold for the removal of the 
social adaptation as well. 
Fig.~\ref{doublecutoff}  gives a comparison of the prevalence with and 
without SA with an upper threshold and a lower threshold. As we increase the SA 
factor $f$; the peak prevalence continues to drop, but a significant drop in the 
peak prevalence is achieved only beyond a critical value of $f$ ($f \approx 4$ in 
the figure). Here, whenever the adaptation factor is large enough to reduce 
the characteristic transmission range to values below its critical value, 
oscillations in prevalence are seen with bigger amplitudes 
lying between the threshold values.

\begin{figure}[!htb]
	  \includegraphics[width=0.95\linewidth]{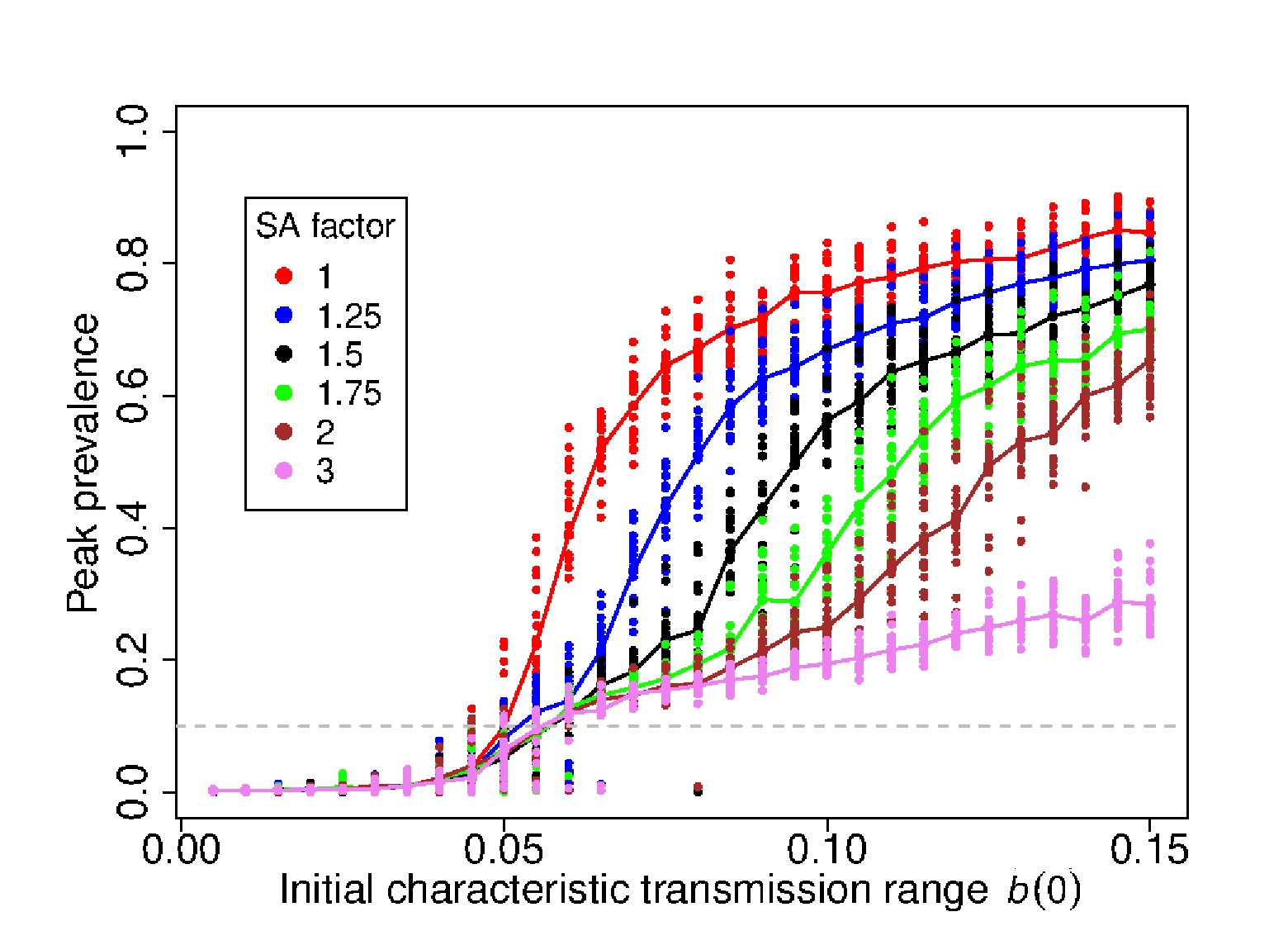}
	\caption{{\bf Simulation results of peak prevalence as a function of initial characteristic transmission range for 
different values of the social adaptation factor $f$.} $i_{c} = 0.1$ and $m_{0} = 0$. The figure shows the influence of varying levels of social adaptation on reducing peak prevalence, with reference to the initial characteristic transmission radius, for a network of spatially static agents.}
	\label{figc01}
\end{figure}

\begin{figure}[!htb]
\includegraphics[width=0.95\linewidth]{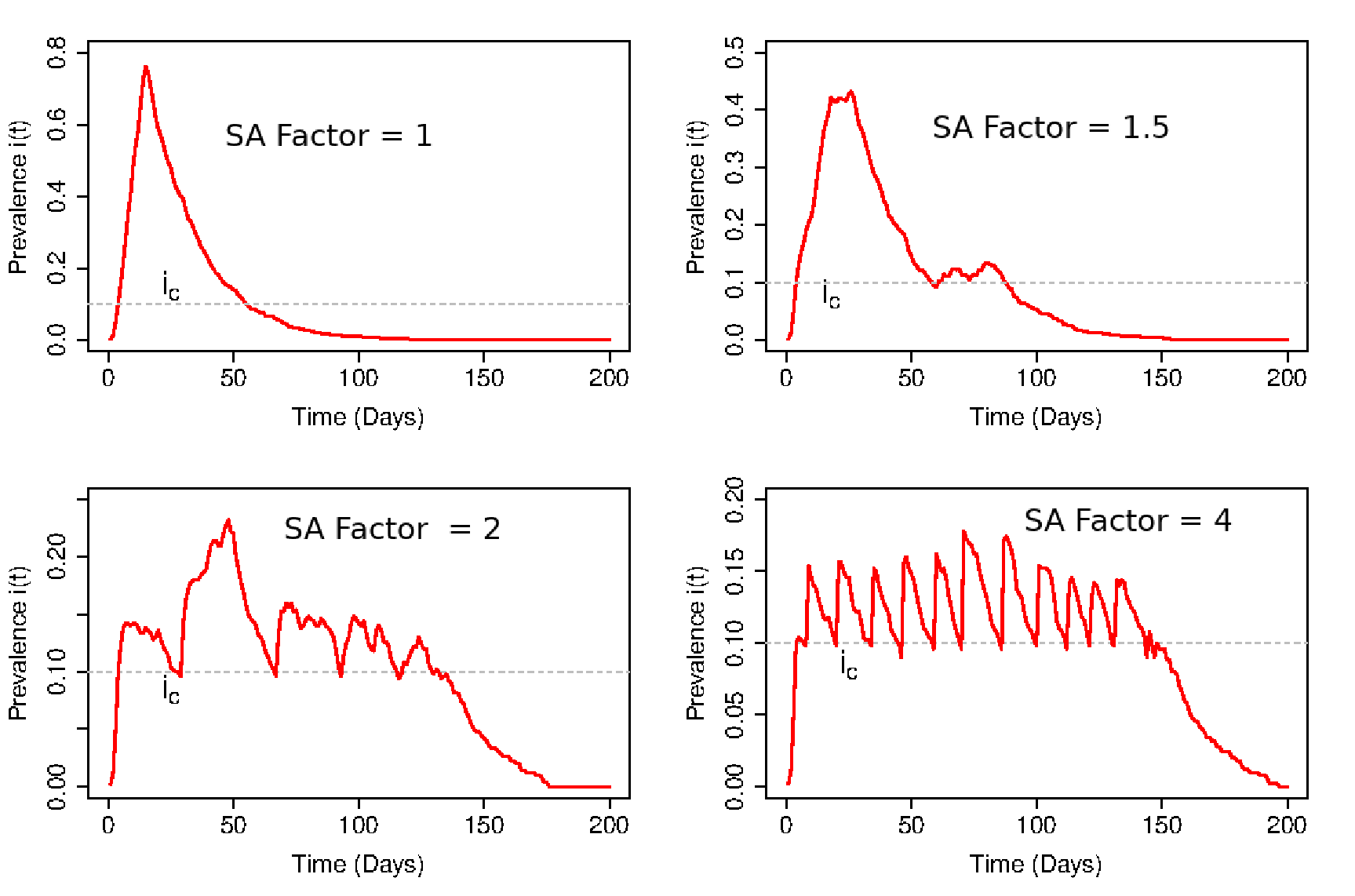}
\caption{{\bf Typical simulation results for variation of prevalence with 
time for different values of the social adaptation factor.}
 $i_c= 0.1$, $m_{0} = 0$, and $b(0) = 0.1$. The figure illustrates the impact of the social adaptation on inducing oscillations in prevalence with a network of spatially static agents.}
	\label{prevlance2}
\end{figure}

When the agents are static, i.e. when $m_{0} = 0$, Fig.~\ref{figc01} gives the 
variation of peak prevalence with the initial characteristic transmission range 
$b(0)$ $i_c = 0.1$. As we increase the SA factor from $1$ to $3$, the peak 
prevalence reduces, but the plateau behavior seen for $m \sim L$ in 
Fig.~\ref{fig8} is less pronounced here. Fig.~\ref{prevlance2} shows the 
prevalence plots for various values of the SA factor. Oscillations in the 
prevalence are seen for higher values of the SA factor. 

\section{Effect of delay in adaptation}
\label{sec4}
So far, we have assumed that the adaptive action by the agents is implemented 
without any delay. So whenever the prevalence crosses the threshold, the agents 
adapt in the very next time step. However, in practice, it is more likely that 
such adaptive action happens with a hold-up due to a delay in the transmission 
of information about the global prevalence or implementation delays. To 
account for such effects, we introduce a delay parameter so that if the 
prevalence goes above or below the threshold in a particular time step,  the 
adaptive action by the agents happens only after a delay of $d$ time steps. 
We can imagine that such a delay can play a significant role in deciding the 
outcome of any attempt to control an epidemic. For highly contagious 
diseases, this delay can lead to situations where the infection has 
already affected a significant fraction of the population even before 
information about global prevalence is available, or any preventive action is 
taken. For a delay of $d$ time steps, we have

\begin{equation}
	b(t+1+d)=\left\{
	\begin{array}{@{}ll@{}}
		\ \frac{b(t)}{f}, & {if}\ i(t)> i_c \label{eqn:eq12}\\		
b(t), & {otherwise}
	\end{array}\right.
\end{equation}

\begin{figure}[!htb] 
	  \includegraphics[width=1\linewidth]{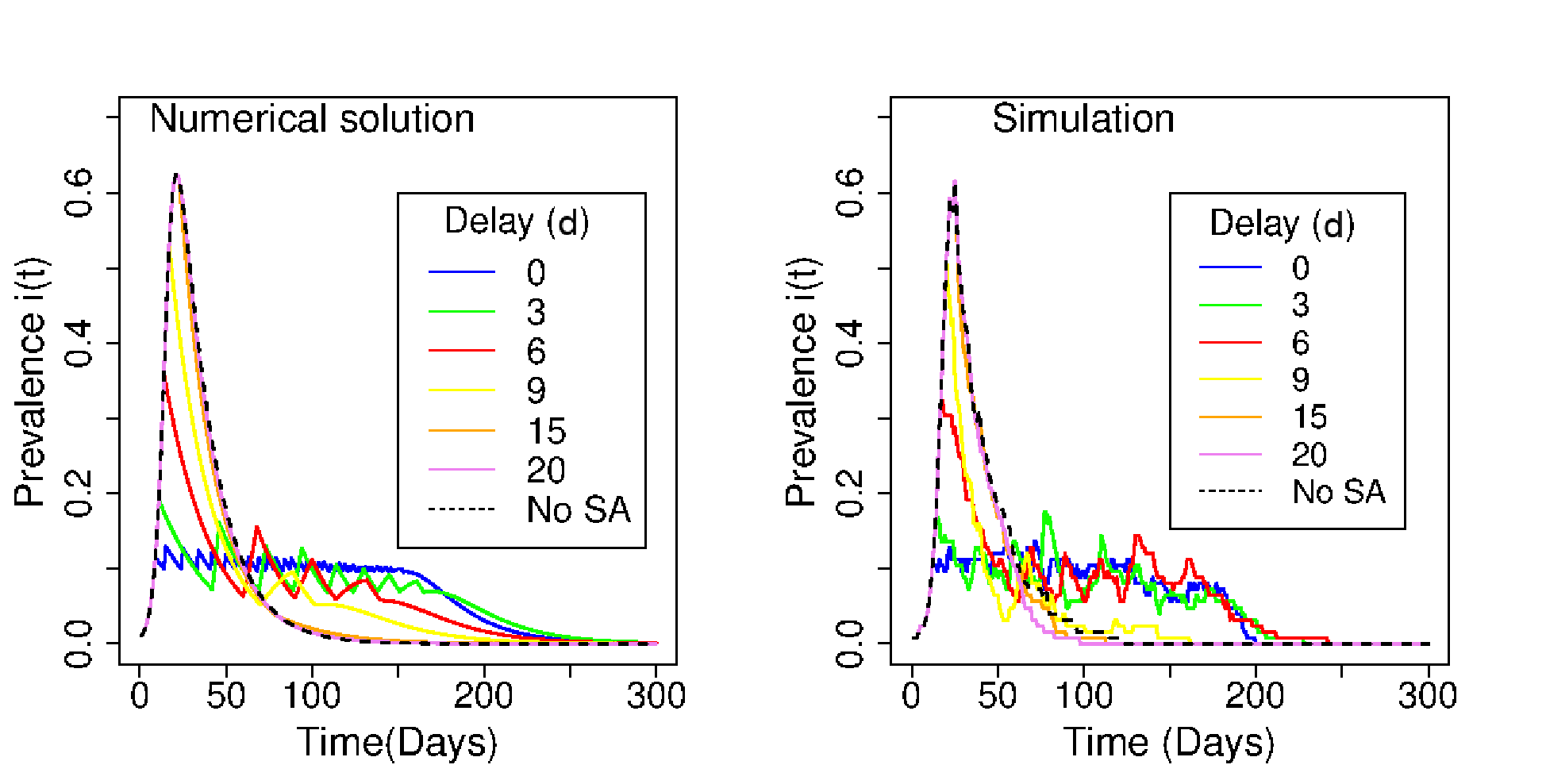}
	\caption{{\bf Variation of prevalence with time when there is a delay of 
$d$ time steps in implementing and removing SA. } Results from numerical 
solution, and simulation are shown. $m_0 \sim L$, $i_c= 0.1$, $b(0) = 0.05$, $\beta = 0.1$, and $\gamma = 0.05$.}
	\label{fig9a}
\end{figure}

\begin{figure}[!htb] 
	  \includegraphics[width=0.95\linewidth]{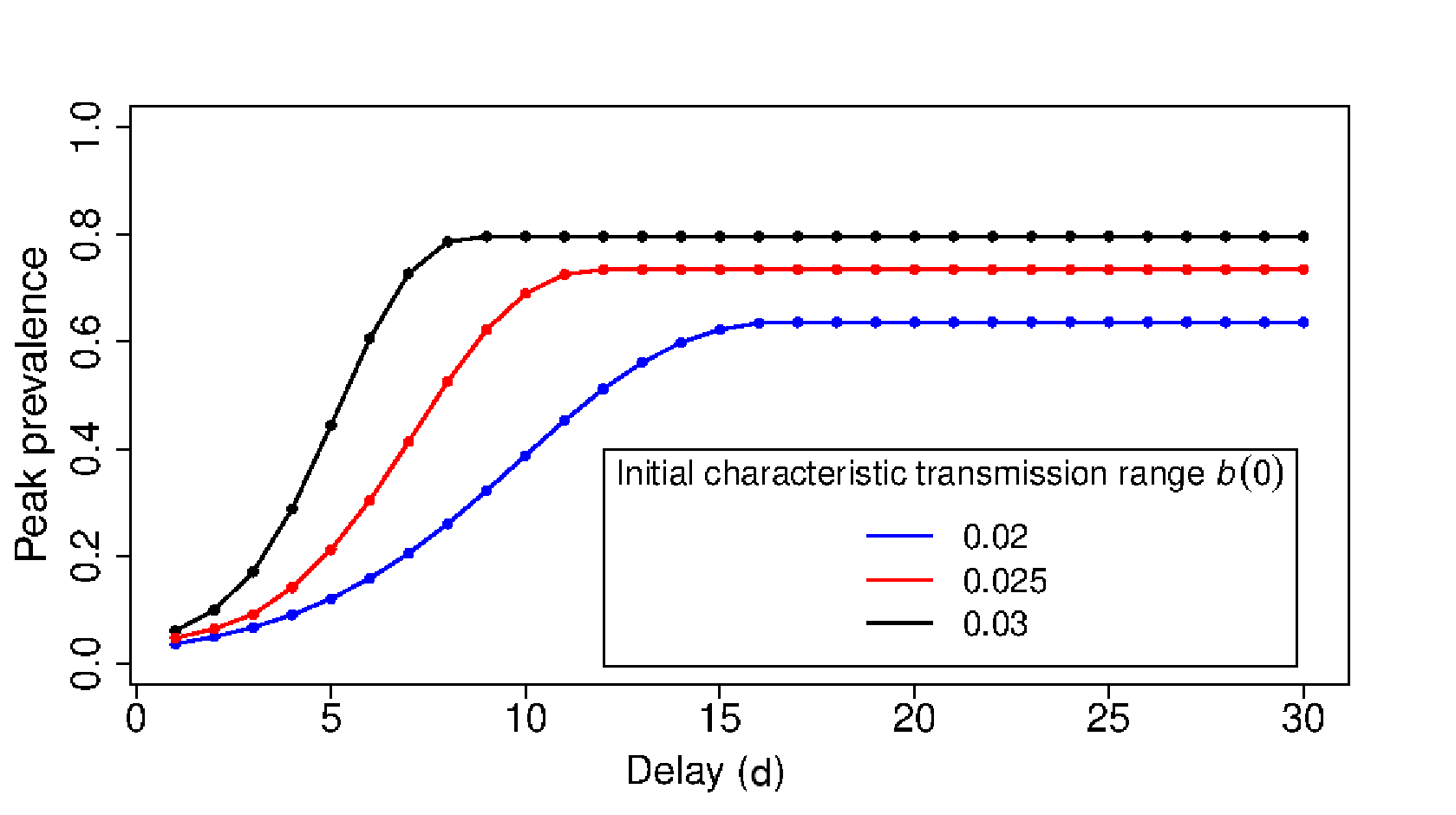}
	\caption{{\bf Variation of peak prevalence with the delay 
parameter $d$ for different values of the initial characteristic transmission range $b(0)$.} $i_c= 0.1$ and $f = 4$, $\beta = 0.5$. This plot emphasizes the importance of implementing social adaptation at the earliest, highlighting that after a specific delay, adaptation cannot contribute to a reduction in peak prevalence.}
	\label{fig10a}
\end{figure}

In Fig.~\ref{fig9a}, we show the numerical and simulation results of prevalence for various values of the delay 
parameter $d$. We can see that as the delay increases, peak prevalence rises significantly, and after a critical value of delay, adaptation becomes 
irrelevant. We can also see that bigger oscillations in the prevalence occur due to the combined effect of social adaptation and the delay. Variation of 
peak prevalence with $d$ for different $b(0)$ is shown in Fig.~\ref{fig10a}. We can see the 
non-linear effect of the delay on peak prevalence, especially for larger values of $b(0)$. This shows the importance of implementing preventive measures with minimum delay, especially for diseases with higher values of transmission range. 

\section{Comparative Analysis with COVID-19 Data}
\label{sec5}
Can the model described above capture any of the features observed in the prevalence data of a real pandemic? We saw that, in the model, adaption can lead to oscillations in the prevalence with varying numbers of peaks and delays between the peaks (Fig. \ref{fig6}), which are features often observed in real-world data of pandemics \cite{Glaubitz2020,Tsuruyma2022}. Here, we attempt a qualitative comparison of our model with data on the COVID-19 pandemic. To this end, we do a comparative analysis of the double threshold model described in Sec.~\ref{sec3}, where agents employ social distancing when the prevalence goes above a particular level and relax the restrictions when the prevalence goes below a level, with publicly available COVID-19 data for different countries \cite{coviddata}. We caution that our aim here is not to fully account for all the observed features of the COVID-19 pandemic but to conduct a qualitative comparison. Several detailed studies exist to predict and compare the real-world prevalence of COVID-19 with epidemiological models \cite{chang2021, Melin2020,Kang2020, bastos2020modeling,giordano2020modelling, lai2020effect}. 

\begin{figure}
    \centering
    \includegraphics[width=13cm]{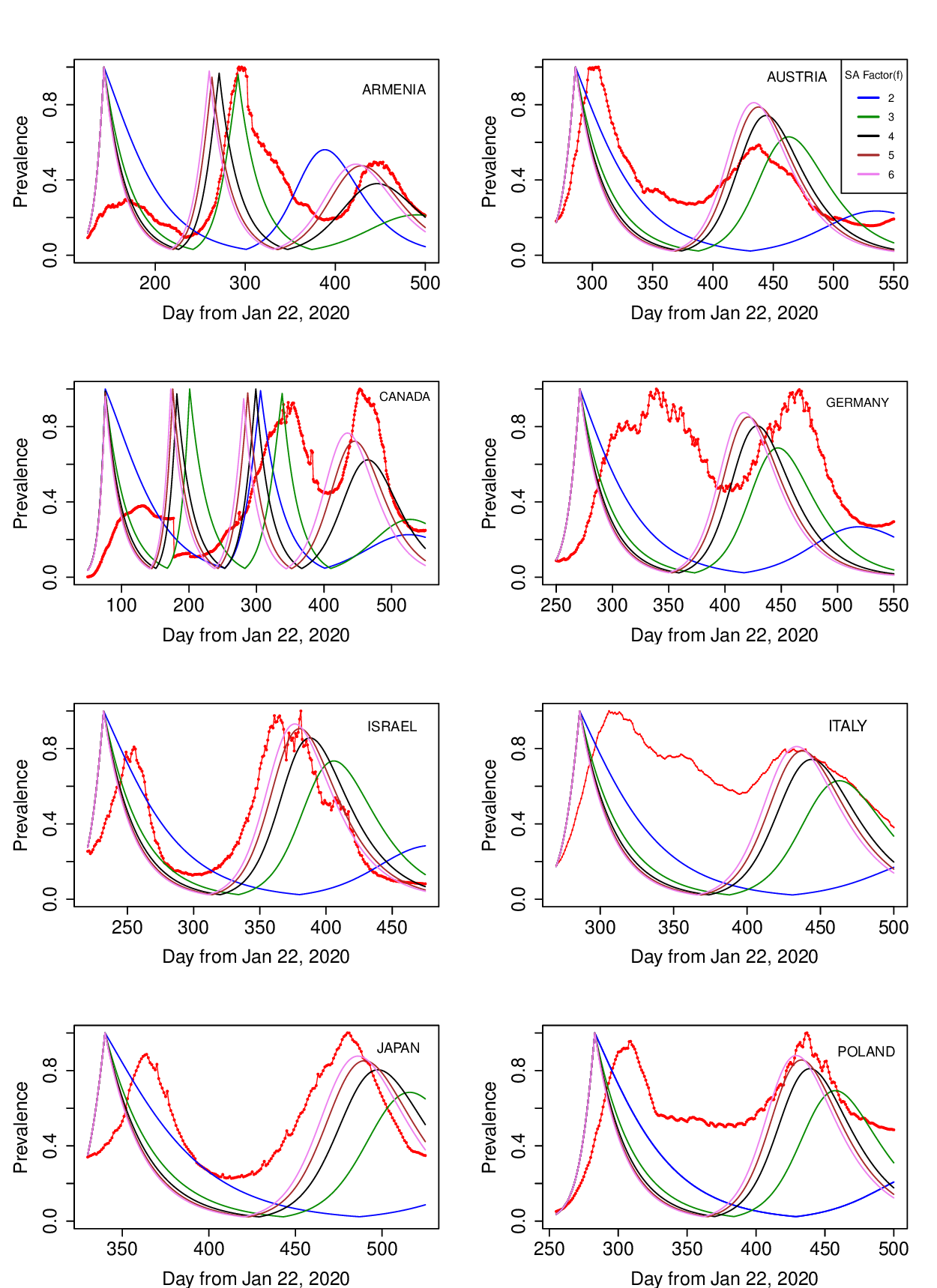}
    \caption{The normalized prevalence over time is calculated utilizing the double threshold model, incorporating an upper threshold of $i_{c2} = 0.2$.  For the lower threshold, we use $i_{c1} = .005$ for all countries except for Armenia and Canada. For them, we use $i_{c1} = .15$ and $0.1$, respectively, so that we will get three distinct peaks for prevalence. The oscillations observed in this model are then contrasted with the fluctuations in normalized actual prevalence data from various countries indicated by red lines. $\rho = 300$, $\beta = 0.4$, $b = 0.02$, $\gamma = 0.05$.}
    \label{fig:fig11}
\end{figure}

We choose the double threshold model (See Fig.~\ref{doublecutoff}) for the comparative study since it captures the real-world scenario of implementing social adaptation at high values of prevalence (corresponding to $i_{c1}$) and relaxing social adaptation at low values of prevalence (corresponding to $i_{c2}$). In Fig.\ref{fig:fig11}, we show the prevalence normalized with respect to its maximum value for eight selected countries, along with the corresponding curves obtained using the double threshold model for various SA factors. Here, we have selected countries that show prominent oscillations in the prevalence during the chosen time period, which is $20^{th}$ January $2020$ to $4^{th} August 2021$. In all cases, we chose $i_{c2} = 0.2$.  We chose $i_{c1} = .005$ for all countries except for Armenia and Canada. For them, we use $i_{c1} = .15$ and $0.1$, respectively, so that we will get three distinct peaks for prevalence. The population density is assumed to be 300 people per unit area. The choice of a transmission probability value is set at $0.4$ in alignment with findings, which indicate that the likelihood of COVID-19 transmission for an unmasked individual ranges from 0.6 without a mask to 0.2 with a mask in close proximity \cite{Agarwal2021}. We fix the value of the initial characteristic transmission radius at $b = 0.02$. From the figure, we can see that, although the numerical match is not great, for specific values of parameters used, the simulated prevalence curves make patterns similar to those of the real-world prevalence data. Specifically, the value of Social Adaption factor $f$, together with other parameters, can give rise to oscillations with varying peak values and different distances between the peaks, which are features observed in many real-world data \cite{coviddata}. Although the model can capture some features qualitatively, we note that it is not expected to capture all the complexities of a real-world pandemic, like the non-homogeneous density of agents, complex mobility patterns, individual agent behavior, etc, to name a few.

\section{Discussion and Conclusion}
\label{sec6}
Spatial effects like mobility and average interpersonal distance are very 
important in deciding the outcome of epidemic dynamics, as amply shown by our 
recent experience with COVID-19. Several recent works have discussed the 
effects of including the spatial aspects in the dynamic of an 
epidemic with adaptive agents 
\cite{Khazaei2021,Liu2021,Glaubitz2020,Just2018}. In general, we can use 
the framework of Random Geometric Graphs for modeling the spread of an epidemic 
incorporating spatial factors. The mobility of agents and their adaptation make the 
graphs evolve in time. In this work, we extended such 
models and considered 
agents who sense the global prevalence of the epidemic and take adaptive 
measures. Agents follow and discard social adaptation based on predefined 
prevalence thresholds. Our results show that such adaptation can have a 
significant effect on the trajectory of the epidemic dynamics. We characterize 
how different levels of adaptation by the agents affect the prevalence of the 
disease and the peak level of infection. Oscillatory prevalence is seen for a 
range of values of the adaptation parameter $f$.  Our results also show that a 
delay in implementing the adaptation can have non-linear effects on peak 
prevalence which shows quantitatively that monitoring the global prevalence 
levels accurately is very crucial so that early intervention based on such 
information is possible. In particular, delay in disseminating information and/or 
delay in taking adaptive measures can accentuate oscillatory prevalence. Our 
results show how simple adaptation behavior by the agents can lead to waves 
during an epidemic, even with SIR dynamics in both fully mixed networks
and static networks.

When spatial factors are included, the condition
for an epidemic outbreak can be written as $1 - 
\gamma + \beta \pi b^2 \rho > 1$ where $\rho$ is the density of the population 
and $b$ 
is the characteristic transmission range of disease. This 
helps to differentiate between factors that can be easily attributed to the 
disease itself (like $\gamma$, $\beta$) and factors related to how the 
agents are distributed over space (like the density $\rho$ or the average 
distance between the agents in the population). Since we can control the latter 
via various non-pharmaceutical strategies like social distancing, mask-wearing, 
partial or complete lock-down etc, the condition thus helps us to clearly 
define the target criteria in order to contain the propagation of disease.

It is quite evident that the behavioral response of agents and authorities can have a huge role in determining the exact course an epidemic will take in a population. In this work, we considered a rather simple social adaptation mechanism within a population facing an epidemic, based on the global prevalence. It is quite conceivable that various stakeholders depend on information different from the global prevalence to decide their course of action. The adaptive mechanism can also change during the course of a pandemic. In the current model, the constant value of the adaptive threshold ensures that successive peaks in the prevalence are reduced.  This makes, for example, the second peak not surpass or equal the maximum value of the initial peak (See Fig.~\ref{fig:fig11}). This contrasts with real-world data of many countries and regions where successive peaks may exceed the preceding ones. Also, we assumed that the adaptation of all agents happens simultaneously and instantaneously, leading to sharp rise or fall in the prevalence curves (See, for example, Fig.~\ref{doublecutoff}). This may be rectified by incorporating a gradual implementation of social adaptation. 

In this work, we considered extreme scenarios where the adaptive agents are fully mobile or not mobile. We can easily extend the 
setting to consider situations where the mixing of agents is more gradual, 
limited spatially, and/or follows a specific pattern. A more realistic setting may be the one in which several 
patches of individuals are connected together by a few long-range connections 
with full mixing within each patch \cite{cornes2022}. We may also introduce heterogeneity 
in the population by considering distributions for parameters characterizing  
social adaptation, prevalence threshold, and mobility \cite{ellison2020, Tri2021, Goel2021}. This 
is especially 
relevant for mobility as infected individuals will, in general, be less mobile. Real-world agents are often distributed heterogeneously in space, and including this aspect in contrast to the homogeneous spatial distribution assumed here will make the model more realistic.
Another obvious direction for future work is to consider the role of spatial 
adaptation in other models of epidemics like SEIR. 
Finally, it will be interesting to look at the effect of social adaptation based 
on local information about the epidemic rather than the global one, as considered 
in the present work. Going further, strategizing agents may be 
considered who will try to optimize individual adaptive actions based on 
information about the prevalence and the action of other agents \cite{Sharma2019}. We 
will explore some of these avenues in a future work.\\

\section*{Code Availability}

The source code for our project is available on our GitHub repository.\\ \href{https://github.com/panickernetworks/SOCIALADAPTATION}{\textcolor{blue}{https://github.com/panickernetworks/SOCIALADAPTATION}}.

\section*{Acknowledgments}
VS acknowledges support from the University Grants Commission-BSR Start-up
Grant No:F.30- 415/2018(BSR).


\begin{thebibliography}{10}

\bibitem{blackdeath}
Austin Alchon, Suzanne.
\newblock {A pest in the land: new world epidemics in a global perspective}.
\newblock University of New Mexico Press 2003;21:ISBN 0-8263-2871-7.  	

\bibitem{coronadata}
\newblock {Coronavirus database}.
\newblock Accessed 1 Aug 2022; https://www.worldometers.info/coronavirus/

\bibitem{Ahmed2021}
Danish A. Ahmed, Ali R. Ansari, Mudassar Imran, Kamal Dingle, Michael B. Bonsall.
\newblock {Mechanistic modelling of COVID-19 and the impact of lockdowns on a short-time scale}.
\newblock Plos one 2021;10(16):1-20.

\bibitem{Gatto2020}
Marino Gatto, Enrico Bertuzzo, Lorenzo Mari, Andrea Rinaldo
\newblock {Spread and dynamics of the Covid-19 epidemic in Italy: Effects of 
emergency containment measures}.
\newblock Proc. Natl. Acad. Sci. USA 2020;117:10484-10491.

\bibitem{Lang2018}
John C Lang, Hans De Sterck, Jamieson L Kaiser, Joel C Miller.
\newblock {{A}nalytic models for SIR disease spread on random spatial networks}.
\newblock Journal of Complex Networks 2018;6(6):948--970.

\bibitem{Colizza2007}
Vittoria Colizza, Alain Barrat, Marc Barthelemy, Alain-Jacques Valleron, Alessandro Vespignani.
\newblock {Modeling the worldwide spread of pandemic influenza: Baseline case 
and containment interventions}.
\newblock PLoS Med 2007;117:4:e13.

\bibitem{Ottar2018}
Ottar N Bjornstad.
\newblock Epidemics, Models and Data using R.
\newblock Springer International Publishing 2018.

\bibitem{Keeling2008}
Matt J. Keeling and Pejman Rohani 
\newblock {Modeling infectious diseases in humans and animals}.
\newblock Princeton University Press 2008.

\bibitem{Rothman2008}
Kenneth J Rothman et al.
\newblock {Modern epidemiology}.
\newblock Wolters Kluwer Health/Lippincott Williams \& Wilkins Philadelphia 2008;3.

\bibitem{Turner2020}
Stefan Thurner, Peter Klimek, Rudolf Hanel.
\newblock {A network-based explanation of why most covid-19 infection curves are 
linear}.
\newblock Proc. Natl. Acad. Sci. USA 2020;117:22684-22689.

\bibitem{Pastor2015}
Romualdo Pastor-Satorras, Claudio Castellano, Piet Van Mieghem, Alessandro Vespignani.
\newblock {Epidemic processes in complex networks}.
\newblock Rev. Mod. Phys 2015;87:925.

\bibitem{Cohen2010}
Cohen, Reuven and Havlin, Shlomo.
\newblock {Complex networks: structure, robustness and function}.
\newblock Cambridge university press 2010.

\bibitem{Junfen22}
Junfeng Fan, Qian Yin, Chengyi Xia, Matjaž Perc.
\newblock {Epidemics on multilayer simplicial complexes}.
\newblock Proc. R. Soc. A 2022;478:20220059.

\bibitem{Sun2022}
Qingyi Sun, Zhishuang Wang, Dawei Zhao, Chengyi Xia, Matjaž Perc.
\newblock{Diffusion of resources and their impact on epidemic spreading in multilayer networks with simplicial complexes}.
\newblock Chaos: Solitons \& Fractals 2022;164:112734.

\bibitem{Shuofan2023} 
Shuofan Zhang, Dawei Zhao, Chengyi Xia, Jun Tanimoto.  
\newblock{Impact of simplicial complexes on epidemic spreading in partially mapping activity-driven multiplex networks}
\newblock Chaos 2023;33(6):063128.

\bibitem{Barthelemy2011}
Marc Barth{\'e}lemy.
\newblock {Spatial networks}.
\newblock Physics Reports 2011;499(1 --3):1-- 101.

\bibitem{Penrose2007}
Mathew Penrose.
\newblock {Random Geometric Graphs}.
\newblock Oxford Scholarship Online 2003;ISBN-13:9780198506263.

\bibitem{chang2021}
Serina Chang et al.
\newblock {{M}obility network models of COVID-19 explain inequities and inform 
reopening}.
\newblock Nature 2021;589:7840:82--87. 

\bibitem{loring2020}
Loring J Thomas et al.
\newblock {{S}patial heterogeneity can lead to substantial local variations in COVID-19 timing and severity}.
\newblock Proc. Natl. Acad. Sci. U.S.A 2020;117:24180--24187. 

\bibitem{wong2020}
 David W. S. Wong, Yun Li.
\newblock {{S}preading of COVID-19: Density matters}.
\newblock Plos one 2020;15:12:e0242398. 


\bibitem{pujari2020}
Bhalchandra S. Pujari, Snehal Shekatkar.
\newblock {{M}ulti-city modeling of epidemics using spatial networks: Application to 2019-nCov (COVID-19) coronavirus in India}.
\newblock Medrxiv 2020.


\bibitem{Melin2020}
Melin P, Monica JC, Sanchez D, Castillo O.
\newblock {Analysis of spatial spread relationships of coronavirus (COVID-19) pandemic in the world using self organizing maps}.
\newblock Chaos, Solitons \& Fractals 2020;138:109914.

\bibitem{Kang2020}
Dayun Kang, Hyunho Choi, Jong-Hun Kim, Jungsoon Choi.
\newblock {Spatial epidemic dynamics of the COVID-19 outbreak in China}.
\newblock International Journal of Infectious Diseases 2020;94:96 -- 102.

\bibitem{WangStati}
Zhen Wang et al.
\newblock{Statistical physics of vaccination}.
\newblock Physics Reports 2016;664:1-113.

\bibitem{Markosocial}
Marko Jusup et al.
\newblock{Social physics}
\newblock Physics Reports 2022;948:1 - 148.

\bibitem{Paulo2022}
Paulo Cesar Ventura, Alberto Aleta, Francisco A. Rodrigues, Yamir Moreno.
\newblock {Epidemic spreading in populations of mobile agents with adaptive 
behavioral response}.
\newblock Chaos, Solitons and Fractals 2022;156:111849.

\bibitem{Arthur2021}
Ronan F. Arthur, James H. Jones, Matthew H. Bonds, Yoav Ram, Marcus W. Feldman.
\newblock {Adaptive social contact rates induce complex dynamics during 
epidemics}.
\newblock  PLOS Computational Biology 2021;17(2):e1008639.

\bibitem{Havlin2020}
Bnaya Gross, Shlomo Havlin.
\newblock {Epidemic spreading and control strategies in spatial modular 
network}.
\newblock Applied Network Science 2020;5(1):95.

\bibitem{Lopez2020}
Leonardo Lopez, Xavier Rodo.
\newblock {The end of social confinement and COVID-19 re-emergence risk}.
\newblock Nature Human Behaviour 2020;4(7):746-755.

\bibitem{Vrugt2020}
Michael te Vrugt, Jens Bickmann, Raphael Wittkowsk.
\newblock {Effects of social distancing and isolation on epidemic spreading 
modeled via dynamical density functional theory}.
\newblock Nat. Commun 2020;11:5576.

\bibitem{Maier2020}
Benjamin F Maier, Dirk Brockmann.
\newblock {Effective containment explains sub exponential growth in recent 
confirmed covid-19 cases in china}.
\newblock Science 2020;368:742-746.

\bibitem{Eli2011}
Eli P. Fenichel et al.
\newblock {Adaptive Human Behavior in Epidemiological Models}.
\newblock Proc. Natl. Acad. Sci. USA 2011;03:108:6306-6311.

\bibitem{Funk2009}
Sebastian Funk, Erez Gilad, Chris Watkins, Vincent A. A. Jansen.
\newblock {The spread of awareness and its impact on epidemic outbreaks}.
\newblock Proc. Natl. Acad. Sci. USA 2009;106:16:6872 - 6877.


\bibitem{Caley2008}
Peter Caley, David J Philp, Kevin McCracken.
\newblock {Quantifying social distancing arising from pandemic influenza}.
\newblock Journal of the Royal Society Interface 2008;5:23:631-9.

\bibitem{Mgosac21}
Marko Gosak, Maja Duh, Rene Markovic, Matjaz Perc. 
\newblock{Community lockdowns in social networks hardly mitigate epidemic spreading.}
\newblock New J. Phys 2021;23:043039.

\bibitem{Khazaei2021}
Hossein Khazaei, Keith Paarporn, Alfredo Garcia, Ceyhun Eksin.
\newblock {Disease spread coupled with evolutionary social distancing dynamics 
can lead to growing oscillations}.
\newblock 2021 60th IEEE Conference on Decision and Control (CDC) 2021;4280-4286.

\bibitem{Jianping2021}
Jianping Huang et al.
\newblock {The oscillation-outbreaks characteristic of the COVID-19 pandemic}.
\newblock National Science Review 2021;8(8):nwab100.

\bibitem{Liu2021}
Haoyan Liu, Xin Wang, Longzhao Liu, Zhoujun Li.
\newblock {Co-evolutionary Game Dynamics of Competitive Cognitions and Public 
Opinion Environment}.
\newblock Frontiers in Physics 2021;9.

\bibitem{Glaubitz2020}
Glaubitz Alina, Fu Feng.
\newblock {Oscillatory dynamics in the dilemma of social distancing}.
\newblock  Proc. R. Soc. A 2020;67- 78.

\bibitem{Just2018}
Winfried Just, Joan Saldana, Ying Xin 
\newblock {Oscillations in epidemic models with spread of awareness}.
\newblock J Math Biol 2018;76:4:1027-1057.


\bibitem{Peng2019}
Xiao-Long Peng, Ze-Qiong Zhang, Junyuan Yang, Zhen Jin.
\newblock{An SIS epidemic model with vaccination in a dynamical contact network of mobile individuals with heterogeneous spatial constraints}.
Commun. Nonlinear Sci. Numer. Simulat 2019;73:52-73.


\bibitem{Buscarino2008}
Arturo Buscarino, Luigi Fortuna, Mattia Frasca, Vito Latora.
\newblock{Disease spreading in populations of moving agents}.
Europhysics Letters 2008;82(3):38002.

\bibitem{Mertens2012}
Stephan Mertens, Cristopher Moore.
\newblock{Continuum percolation thresholds in two dimensions}.
Phys. Rev. E 2012;86(6):061109.


\bibitem{Tsuruyma2022}
Tatsuaki Tsuruyama.
\newblock{Nonlinear model of infection wavy oscillation of COVID-19 in Japan based on diffusion kinetics.}
\newblock Sci Rep 2022;12:19177.


\bibitem{coviddata}
\newblock{Datahub database. COVID-19 country-wise dataset from 22 January 2020 to 4 August 2021.}
\newblock \href{https://datahub.io/core/covid-19}{https://datahub.io/core/covid-19}.

\bibitem{bastos2020modeling}
Saulo B. Bastos, Daniel O. Cajueiro
\newblock{Modeling and forecasting the early evolution of the Covid-19 pandemic in Brazil}.
\newblock Scientific Reports 2020;10(1):19457.


\bibitem{giordano2020modelling}
Giulia Giordano et al.
\newblock{Modelling the COVID-19 epidemic and implementation of population-wide interventions in Italy}
\newblock Nature medicine 2020;26:6:855--560.

\bibitem{lai2020effect}
Shengjie Lai et al.
\newblock{Effect of non-pharmaceutical interventions to contain COVID-19 in China}.
\newblock Nature 2020;585:7825:410--413.

\bibitem{Agarwal2021}
Amit Agrawal,  Rajneesh Bhardwaj.
\newblock{Probability of COVID-19 infection by cough of a normal person and a super-spreader}
\newblock Phys Fluids (1994) 2021;33(3):031704.

\bibitem{cornes2022}
Fernando E. Cornes, Guillermo A. Frank, Claudio O. Dorso.
\newblock {COVID-19 spreading under containment actions}.
\newblock Physica A 2022;588:126566.

\bibitem{ellison2020}
G. Ellison.
\newblock {Implications of heterogeneous SIR models for analyses of 
COVID-19}.
\newblock tech. rep., National Bureau of Economic Research 2020.


\bibitem{Tri2021}
S. Triambak and D.P. Mahapatra.
\newblock {A random walk Monte Carlo simulation study of COVID-19-like infection spread}.
\newblock Physica A: Statistical Mechanics and its Applications 2021;9.

\bibitem{Goel2021}
Rahul Goel, Loïc Bonnetain, Rajesh Sharma, Angelo Furno.
\newblock {Mobility-based SIR model for complex networks: with case study Of COVID-19}.
\newblock Soc. Netw. Anal. Min 2021;11:105.


\bibitem{Sharma2019}
Anupama Sharma, Shakti N. Menon, V. Sasidevan, Sitabhra Sinha.
\newblock {Epidemic prevalence information on social networks can mediate 
emergent collective outcomes in voluntary vaccine schemes}.
\newblock PLOS Computational Biology 2019;15:5.

\end{thebibliography}
\end{document}